\documentclass[10pt]{iopart}
\usepackage{comment}
\usepackage{gensymb}
\usepackage{textcomp}
\usepackage{float}
\usepackage{cite}
\usepackage{braket}
\usepackage{hyperref}
\usepackage{caption}
\usepackage{url}
\usepackage{rotating}
\usepackage{float}
\expandafter\let\csname equation*\endcsname\relax
  \expandafter\let\csname endequation*\endcsname\relax
\usepackage{amsthm}
\usepackage{amsmath}
\usepackage{amssymb}
\usepackage{graphicx}				
\usepackage{array}					
\usepackage{color}							
\usepackage{amsfonts}
\usepackage{float}

\usepackage {listings}

\begin{document}

\title{Optical nanofibres and neutral atoms}
\author{Thomas Nieddu, Vandna Gokhroo \footnote{Current address: Dept. of Physics and Astronomy, Washington State University, Pullman, WA-99163, USA} and S\'{\i}le Nic Chormaic}
\address{Light-Matter Interactions Unit, Okinawa Institute of Science and Technology Graduate University, Onna-son, Okinawa 904-0495 Japan}

\ead{sile.nicchormaic@oist.jp}
\begin{abstract}
Optical nanofibres are increasingly being  used in cold atom experiments due to their versatility and the clear advantages they have when developing all-fibred systems for quantum technologies. They provide researchers with a method of overcoming the Rayleigh range for achieving high intensities in a focussed beam over a relatively long distance, and can act as a noninvasive tool for probing cold atoms. In this review article, we will briefly introduce the theory of mode propagation in an ultrathin optical fibre and highlight some of the more significant theoretical and experimental progresses to date, including the early work on atom probing, manipulation and trapping, the study of atom-dielectric surface interactions, and the more recent observation of nanofibre-mediated nonlinear optics phenomena in  atomic media. The functionality of optical nanofibres in relation to the realisation of atom-photon hybrid quantum systems is also becoming more evident as some of the earlier technical challenges are surpassed and, recently, several schemes to implement optical memories have been proposed. We also discuss some possible directions where this research field may head, in particular in relation to the use of optical nanofibres that can support higher-order modes with an associated orbital angular momentum.
\end{abstract}
\pacs{37.10.Gh, 42.50.-p}
\submitto{\JOPT}
\maketitle
\ioptwocol

\section{Introduction}
\hspace{0.5cm} The development of quantum-based devices is becoming increasingly necessary for future technological advances and, as a result, enormous effort is being made to find suitable experimental platforms, one of which is cold, neutral atoms.  Strong coupling can be viewed as an essential requirement to devise quantum networks based on neutral atoms \cite{Cirac97,Boozer07,Kimble08} and, until recently, free-space cavities were the most promising method for enabling an atomic system to enter this regime \cite{Miller05}.   Very recently, Kato and Aoki \cite{Aoki2015} demonstrated strong coupling for a single trapped atom and an all-fibred optical cavity based on an optical nanofibre (ONF) spliced between two fibre Bragg grating mirrors, thereby illustrating the potential offered by ONFs for future quantum technological advances.\\

Another distinct advantage of optical nanofibres is that they facilitate the study of nonlinear phenomena in an atomic medium without requiring a tightly focussed laser beam, as in free-space optics.  If light is to be tightly focussed, the interaction distance is essentially limited by the Rayleigh length, $L_{\text{R}}$, given by $L_{\text{R}} = \pi {w_0}^2 / \lambda$, where $w_0$ is the beam radius at the narrowest point and $\lambda$ is the laser wavelength. If we consider a 780 nm beam, corresponding to the D$_2$ line in $^{87}$Rb, the interaction length is limited to about 4 $\mu$m if the beam is focussed to a 1 $\mu$m radius.  ONFs allow us to overcome this limitation \cite{Tong04}, due to the strong evanescent field at their waist (i.e. the thinnest region of the nanofibre) which can contain most of the power injected into them \cite{LeKien2004}.  This evanescent field may interact with atoms in the vicinity of the fibre \cite{Lekien06} and spontaneously emitted light from excited atoms can preferentially couple into the nanofibre-guided modes \cite{Lekien05,Masalov13}. Such properties have led, in recent years, to an explosion of interest in integrating ONFs into cold atomic systems in order to fully exploit the advantages that they offer.\\

In this review paper, we present a brief introduction to mode propagation in ONFs and provide the reader with pertinent references  in relation to the fabrication of such fibres, whether for fundamental mode or higher order mode (HOM) propagation.  Next, we summarise some of the earlier ONF work with atomic vapours and cold atoms, leading up to the realisation of ONF-based cold atom traps and the achievement of strong coupling between a single atom and an ONF-based cavity.  Very recent work on nonlinear optics phenomena, such as electromagnetically induced transparency (EIT) and slow light using ONFs with cold atoms, is then discussed. Finally, we aim to provide some insight into possible future directions of the field.  Note that, depending on the wavelength being used, HOM propagation may require the fibre waist to be larger than a micron, in which case it should be referred to as an optical microfibre (OMF). For clarity, we will assume that all fibres discussed in this review are submicron in diameter, i.e. ONFs, unless explicitly stated otherwise. \\

\section{Mode propagation in an optical fibre}
\subsection{Evanescent field and guided modes of an optical fibre}
\hspace{0.5cm} In the following, we derive expressions that describe the evanescent field and guided modes in a standard optical fibre. An extensive development of the mathematics presented here can be found, for example, in \cite{Yariv1985,Snyder2000}. An optical fibre can be approximated to a perfectly cylindrical, step-index fibre in which the refractive index of the core, $n_{\text{core}}$, is greater than the refractive index of the cladding, $n_{\text{clad}}$. Mathematically, the fibre is represented by the following index profile
\begin{equation}
n(r)=\left\{
                \begin{array}{ll}
                  n_{\text{core}}, \: 0< r \leqslant a \\
                  n_{\text{clad}}, \: a < r \leqslant \rho\\
                \end{array}
              \right.
\end{equation}
where $a$ denotes the core radius, $r$ is the radial position from the fibre centre and $\rho$ is the cladding radius as defined from the centre of the fibre. If we assume the fibre to be isotropic and source-free, Maxwell's equations for light propagating within this system can be reduced to the following set of Helmoltz equations
\begin{equation}
\label{eq:helmoltz}
(\nabla^{2}_{\text{T,cyl}} + n^2k^2-\beta^2)\psi_z=0,
\end{equation}
with $\nabla^2_{\text{T,cyl}} = \nabla^2 - \partial^2/ \partial z^2$ representing the transverse Laplace operator in cylindrical coordinates, $k = 2\pi/\lambda$ is the wave number, $\beta$ is the modal propagation constant and $\psi = \left\{E_z, H_z \right\}$ represents the $z-$component of either the electric or magnetic field of the input light, respectively. This equation is separable into its individual components \cite{Yariv1985} and solutions may take the form
\begin{equation}
\label{eq:field}
\psi_{z}(r,\phi)=R(r)e^{\pm il\phi},
\end{equation}
with $R(r)$ being the radial part and the exponential term representing the azimuthal part. Here $l = 1, 2, 3, ...$, and $z$, $r$ and $\phi$ denote  the longitudinal, radial and azimuthal components of the field , respectively. Inserting this into Eq. \eqref{eq:helmoltz} gives
\begin{equation}
\left[\frac{\partial^2}{\partial r^2}+\frac{1}{r}\frac{\partial}{\partial r}+\left(n^2k^2-\beta ^2 - \frac{l^2}{r^2}\right)\right]R(r)=0,
\end{equation}
which is a Bessel differential equation, the solutions of which are Bessel functions of order $l$. Two cases can be distinguished depending on the sign of $n^2k^2-\beta^2$ and their general solutions can be calculated to yield \cite{Snyder2000}
\begin{equation}
\begin{array}{ll}
R(r)=c_1J_l(hr)+c_2Y_l(hr) \:\: \text{if}\; n^2k^2-\beta^2>0,\\
R(r)=c_3I_l(qr)+c_4K_l(qr) \:\: \text{if}\;n^2k^2-\beta^2<0\\
\end{array}
\end{equation}
where $h=\sqrt{n^2k^2-\beta^2}$, $q=\sqrt{\beta^2-n^2k^2}$, the $c_i$ represent constants, $J_l$ and $Y_l$ are Bessel functions of order $l$ of the first and second kind, respectively, and $I_l$ and $K_l$ are modified Bessel functions of order $l$ of the first and second kind, respectively.\\

For confined propagation of  light at the fibre's waist, the effective refractive index, $\beta / k$, should be larger than the cladding's refractive index, which leads to the case $n^2k^2-\beta^2<0$. For $r\rightarrow\infty$, $I_l$ diverges and leads to nonphysical solutions. Therefore, $c_3=0$ and we are left with the term $c_4K_l(qr)$ in the radial part of the field (see Eq. \eqref{eq:field}) describing the exponential decay of the evanescent field at the waist of the fibre.\\

Solving Eq. \eqref{eq:helmoltz} for the fields in the core yields four eigenvalue equations, corresponding to the four possible guided modes inside a cylindrical optical fibre, namely the HE, EH, TE and TM modes. We list their analytical expressions here, without further development \cite{Yariv1985}: \\
$\text{EH}_{lm}$ modes:
\begin{equation}
\frac{J_{l+1}(ha)}{haJ_l(ha)}= \frac{n_{core}^2+n_{clad}^2}{2n_{core}^2}\frac{K'_{l}(qa)}{qaK_l(qa)}+\left(\frac{l}{(ha)^2}-W\right)
\end{equation}
$\text{HE}_{lm}$ modes:
\begin{equation}
\frac{J_{l+1}(ha)}{haJ_l(ha)}= -\left( \frac{n_{core}^2+n_{clad}^2}{2n_{core}^2}\right)\frac{K'_{l}(qa)}{qaK_l(qa)}+\left(\frac{l}{(ha)^2}-W\right)
\end{equation}
$\text{TE}_{0m}$ modes:
\begin{equation}
\frac{J_{1}(ha)}{haJ_0(ha)}= -\frac{K_{1}(qa)}{qaK_0(qa)}
\end{equation}
$\text{TM}_{0m}$ modes:
\begin{equation}
\frac{J_{1}(ha)}{haJ_0(ha)}= -\frac{n_{clad}^2K_{1}(qa)}{qan_{core}^2K_0(qa)}
\end{equation}
\begin{equation}
\begin{split}
\text{with} \;\; W =\Biggl[ & \left(\frac{n_{core}^2 - n_{clad}^2}{2n_{core}^2}\right)^2 \left(\frac{K'_l(qa)}{qaK_l(qa)}\right)^{2} \\& + \left(\frac{l \beta}{n_{core} k_0} \right)^2 \left(\frac{1}{q^2a^2}+\frac{1}{h^2a^2} \right)^2\Biggr]^{1/2}.
\end{split}
\end{equation}

\noindent In these expressions, $K'_{l}$ denotes $dK(qa)/d(qa)$. \\

\subsection{Guided modes of an ONF}
\begin{figure}[h!]
\centering
        \includegraphics[width=0.49\textwidth]{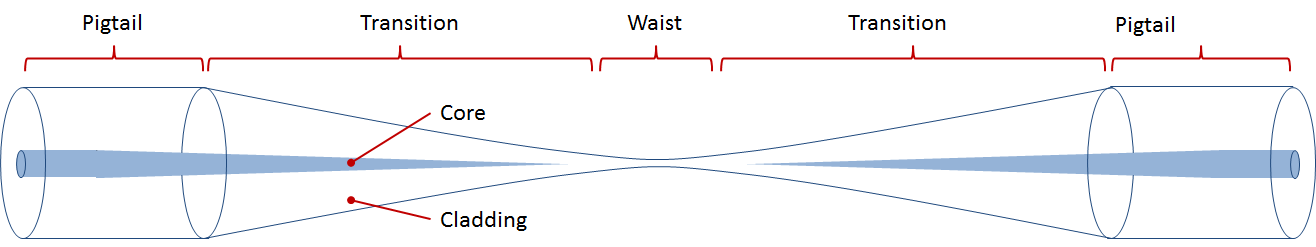}
  \caption{Schematic illustration of an ONF. Three parts can be distinguished; the pigtail, which is an unaltered region of the standard fibre, the taper, where the core size is shrunk until it vanishes, and the waist, where the core has completely vanished, forcing the guided light to create an evanescent field outside the fibre. }
  \label{fig:MNF}
\end{figure}

\hspace{0.5cm}Fabrication of an ONF usually consists of heating  a commercially available optical fibre, stripped of its acrylic coating, to a temperature close to the glass melting point (for example, ~1550$\degree$C for silica). Under these conditions, the fibre is in a plastic regime and can be elongated without breaking. As the fibre is pulled, three regions can be distinguished, as illustrated in Fig.\ref{fig:MNF}.  These are the pigtail, the taper and the waist, which, if small enough, leads to the generation of a significant evanescent field when light is launched into the fibre \cite{Bures99}.

Several fabrication methods have been reported \cite{Tong03,Brambilla04,Ward06,Shi06,Ding10,Tong12,Ward14,Nagai14} and the one used generally depends on one's needs, aims and resources. For cold atom experiments, the so-called \emph{flame-brushing technique} \cite{Brambilla04} is usually preferred since it offers fine control over the taper shape and the waist size, two important parameters for modal selection. Routinely, fibres of the order of 350 nm can be fabricated.  Similar advantages were recently reported using a microheater-based fibre-pulling rig \cite{Heidi14}.\\

\begin{figure}[h!]
\centering
        \includegraphics[width=0.49\textwidth]{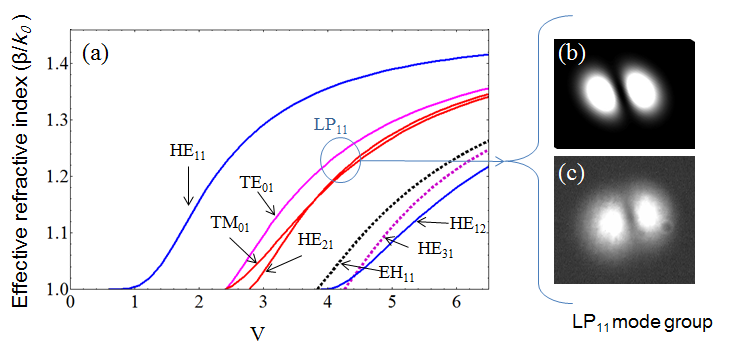}
  \caption{(a) Plot of the effective refractive index versus the V-number showing the modes that can propagate in an optical nanofibre.  For a given effective refractive index, a reduction in the V-number reduces the number of modes permitted. The circle indicates the family of the first four higher order modes, namely TE$_{01}$, TM$_{01}$, HE$_{21,\text{even}}$ and HE$_{21,\text{odd}}$ collectively referred to as the LP$_{11}$ group. (b) Simulated and (c) experimental images of the output of an ONF when light is guided in the LP$_{11}$ group of modes. Reproduced from \cite{Kumar15HOM}.}
  \label{fig:Vnum}
\end{figure}

When light is launched into a fibre, it can couple into several of the available guided modes \cite{Ravets213}. Light propagates differently in each of these modes, a property characterised by the modal propagation constant, $\beta$. This difference in propagation gives rise to modal interference and may result in unwanted beam profiles at the output of the fibre. However, one can limit the number of propagating modes by reducing the dimensionless V-number, obtained from solving Maxwell's equations for an optical fibre \cite{Yariv1985,Snyder2000}. The V-number is determined from
\begin{equation}
\label{eq:Vnumber}
\text{V} = \frac{\pi d}{\lambda} \sqrt{n^2 _{\text{core}} - n^2 _{\text{clading}}},
\end{equation}
where $d=2a$ is the fibre waist diameter, $\lambda$ is the wavelength of light, and $n_{\text{core}}$, $n_{\text{clading}}$ are the refractive indices of the fibre core and cladding, respectively.  As shown in Fig. \ref{fig:Vnum}(a), the modal limitation is easily understood when one plots the effective refractive index, $n_{\text{eff}} = \beta / k$, versus the V-number. For a given wavelength, the V-number and, consequently, the number of allowed modes, can be reduced by decreasing the fibre diameter. This is achieved during the ONF fabrication as this process consists of tapering the radius of a normal optical fibre until it reaches a few hundreds of nanometers in diameter.\\

\begin{figure}
\centering
        \includegraphics[width=0.40\textwidth]{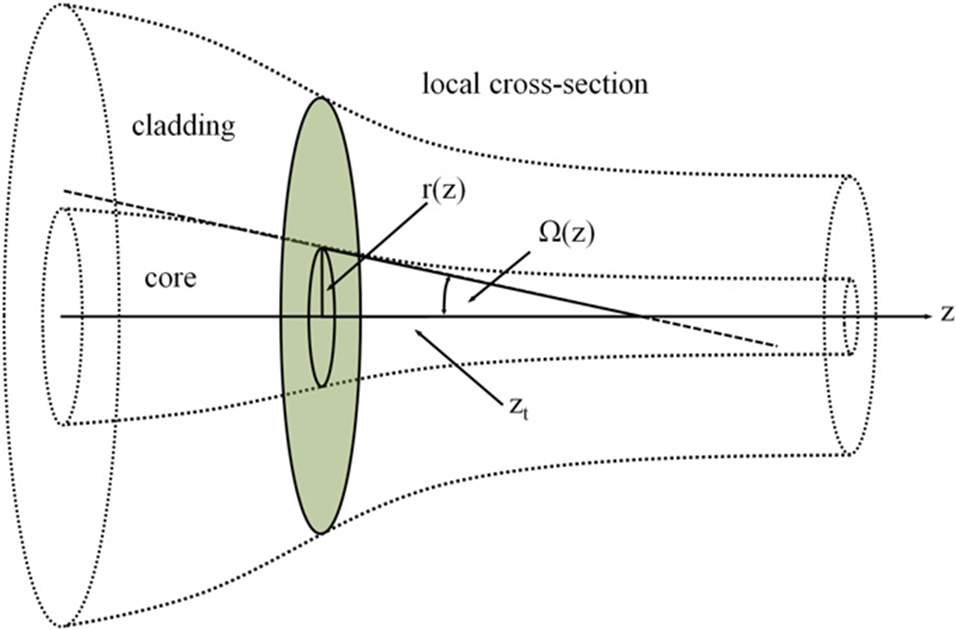}
  \caption{Diagram of the pigtail and transition region, showing the taper angle, $\Omega (z)$, the core radius, $r(z)$, and taper length, $z(t)$. Reproduced from \cite{Frawley2012}.}
  \label{fig:adiab}
\end{figure}

Changing the structure of an optical fibre influences its birefringence properties and may result in losses if not done carefully. These losses can be minimised by ensuring that the adiabaticity criterion is met during fabrication \cite{Love91}. As stated above, light injected into an ONF can couple into different fibre modes, each  yielding a different propagation constant. For example, Fig. \ref{fig:Vnum}(b) shows the simulated  and Fig. \ref{fig:Vnum}(c) the experimental output profiles obtained when one limits the fibre-guided modes to the LP$_{11}$ group by reducing the fibre diameter, thereby reducing the V-number to a value close to 2.771, i.e. the HE$_{21}$ cut-off. The double-lobe pattern is the result of interference between the different submodes of this group. At the start of the taper region, these modes are not in phase and consequently experience beating, which results in losses. The beat length, $z_b = 2\pi / (\beta_1 - \beta_2)$, has  to be kept as short as possible compared to the taper length, $z_t \approx r(z) / \Omega (z)$, where $r(z)$ is the core radius and $\Omega (z)$ is the taper angle (see Fig.\ref{fig:adiab}). In the ideal case $z_b \ll z_t$; this is achieved when the taper angle is as small as possible and, in this case, light at the taper couples entirely to a single mode.\\

\begin{figure*}
\centering
        \includegraphics[width=0.9\textwidth]{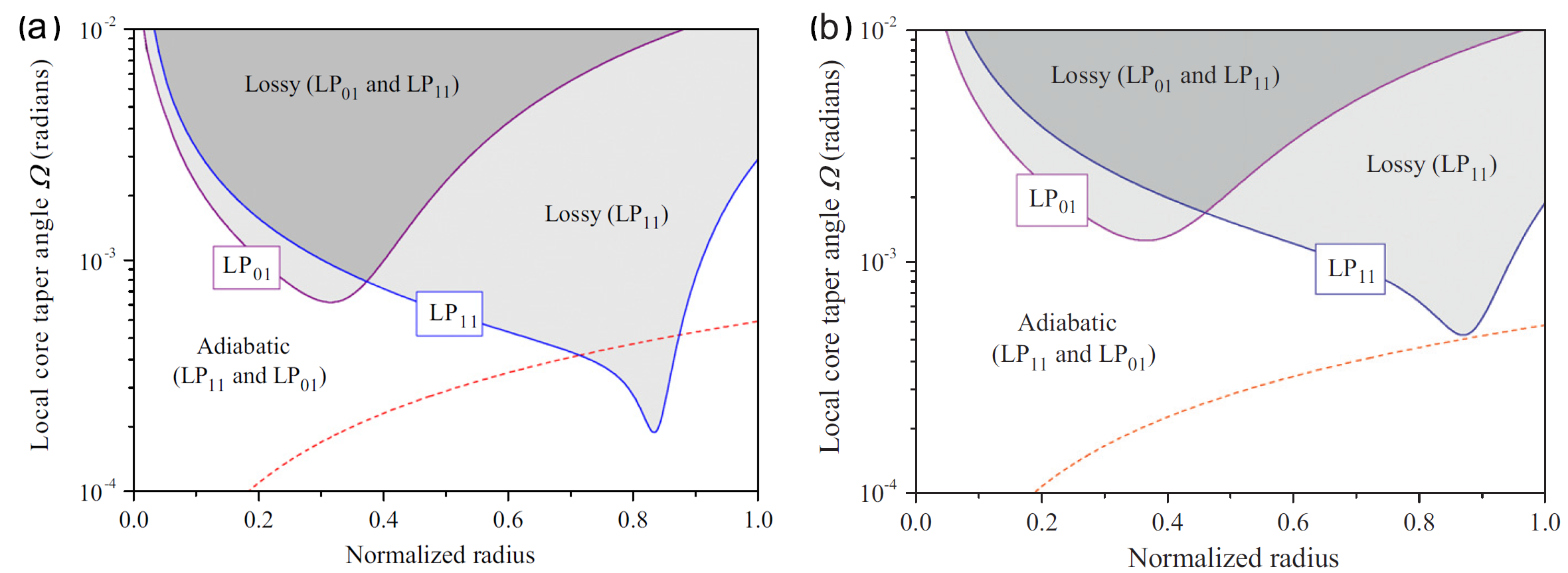}
  \caption{Plots of the adiabatic condition and lossy regimes for the LP$_{01}$ and LP$_{11}$ modes in (a) an exponential taper of a 125 $\mu$m diameter fibre, and (b) an exponential taper of an 80 $\mu$m diameter fibre. Reproduced from \cite{Frawley2012}.}
  \label{fig:125v80}
\end{figure*}

Our group has shown that similar reasoning can be applied to derive the adiabatic criterion for efficient HOM propagation \cite{Colan11}. It is possible to limit propagation of the HE$_{21}$ mode by careful control of the pulling length and using commercially available 80 $\mu$m diameter fibre \cite{Frawley2012}, which has a different cladding-to-core ratio, instead of the usual 125 $\mu$m. As plotted in Fig.\ref{fig:125v80}, reducing the cladding- to-core ratio shifts the adiabatic condition on the local core taper angle. This is a result of the fact that fewer modes can propagate within the cladding, thereby confining them in the core and reducing the beat length. The choice of this ratio is, therefore, crucial for HOM optical nanofibre experiments. The technique was later refined by Ravets \textit{et al.}, but for 50 $\mu$m diameter fibres \cite{Ravets13}. They matched the HOM adiabatic criterion even better and obtained HOM propagation efficiency for the LP$_{11}$ group up to 97.8$\%$. Recently, a heat-and-pull fibre rig for ultra-high transmission of both fundamental mode and HOM fibres was reported by Hoffman \textit{et al.} \cite{Hoffman14}. Using the same fibre-pulling rig, the group reported observation of the spatial evolution of HOMs in the LP$_{11}$ group, over the whole fibre length, via Rayleigh scattering imaging \cite{HoffmanArxiv}. Additionally, they were able to identify and control which submodes of the LP$_{11}$ group were propagating within the OMF.\\

\section{Neutral atoms and optical nanofibres}
\subsection{Probing, sensing and characterising atomic systems with an ONF}

\hspace{0.5cm} ONFs have attracted considerable interest for particle trapping and manipulation and, more recently, for the development of quantum technologies \cite{sensors}. Such fibres are particularly praised for their potential role as efficient and reliable photonic communication channels between distant quantum systems \cite{Lekien05}. Atoms localised close to the ONF's surface can emit light that couples to its guided modes \cite{Nayak07,Russell09,Masalov14}. This, in turn, enhances the photon emission rate \cite{Lekien12}, and can be engineered to generate a lateral translation-invariant force on the atoms \cite{Scheel15}. The coupled photons in these modes are correlated in either a bunching or an anti-bunching fashion, as demonstrated experimentally by Nayak \textit{et al.} \cite{Nayak09}. Due to the enhanced evanescent field produced at the waist, the guided light can interact with surrounding atoms via absorption \cite{Lekien06}. The first experimental demonstration of this property was made in 2007 by Sagu\'{e} \textit{et al.} \cite{Sague07} for spectroscopy purposes. They created a cloud of cold caesium atoms in a magneto-optical trap (MOT) which was superimposed with an ONF. A probe beam was then launched into the ONF and its absorption measured on an avalanche photodiode (APD). Emitted light subsequently couples back into the fibre,  the propagation direction of which is dictated by the polarisation of the initially injected light \cite{LeKien2014} or the atoms' arrangement in the immediate vicinity of the fibre \cite{LeKien22014}.\\

Extensive work on in-fibre spectroscopy of hot vapours \cite{Spillane08,Hendrickson09,Hendrickson10,Jones14} was also carried out in parallel. However, introduction of an ONF in an atomic vapour often results in the atoms sticking to the ONF's surface, thereby reducing its efficiency as a probing tool. Several ways of dealing with this problem have been explored by the UMBC group. For example, they demonstrated the desorption of rubidium atoms by launching low power, non-resonant light in the ONF \cite{Hendrickson09}, or by placing it on a flange-mounted heater \cite{Lai13}. They also proposed to switch to a vapour of metastable xenon atoms\cite{Pittman13}, since it is a noble gas and consequently does not interact as much with the fibre. They showed  saturated absorption at the nW level in this system. The same group recently reported the observation of a nonlinear optical effect, EIT, in a rubidium vapour \cite{Jones15}. \\

\hspace{0.5cm}Due to the ultra-low pressures needed in cold atom experiments in contrast to vapour experiments, the problem of atom adsorption on the nanofibre  does not occur to a detrimental level. This makes the ONF an excellent tool to probe and characterise atom clouds created in a MOT \cite{Morrissey09,Deasy10}. Several experiments carried out by Russell \textit{et al.} \cite{Russell2012,Russell2013,Russell2014} used an ONF embedded in such a system to measure the temperature of a cloud of $^{85}$Rb atoms using different techniques. In \cite{Russell2013}, temperature variations of the order of $\mu$K created by changes in the cooling-lasers' alignment, intensities or detuning, were detected by comparing the fluorescence signals coupling into the ONF throughout the different stages of a release-recapture process. A dark-MOT \cite{Russell2014} of the same atomic species was also characterised (via the atom density and loading time) using similar methods. More recently, Grover \textit{et al.} \cite{Grover2015} performed temperature measurements in a MOT, but this time on $^{87}$Rb atoms. The characterisation method, although relying on fluorescence collected from an ONF, used the intensity-intensity correlation function to extract the velocity of atoms surrounding the fibre and, hence, their temperature. It has also been proposed  to use an ONF as position sensors of trapped atoms in an optical lattice \cite{Hennessy14}.\\

\subsection{Fibre-based traps for laser-cooled atoms}

The interaction of atoms with an evanescent light field has long been known \cite{Cook82,Balykin88,Hajnal89,Feron93,Feron94}.  In the 1990's, it was proposed \cite{Ol'Shanii93}, and later demonstrated \cite{Renn95}, that atoms could be guided inside a hollow-core fibre using the optical dipole force resulting from the interaction with the guided light. In these experiments, a laser beam, red-detuned from the  average transition frequency of the D$_2$ spectroscopic lines for $^{85}$Rb and $^{87}$Rb, was launched inside the hollow core fibre. The red-detuned light generated an attractive dipole force for atoms and they were guided inside the fibre. Later, Ito \textit{et al.}\cite{Ito96} designed a similar experiment using blue-detuned light to repel the atoms from the fibre's inner surface and guide them along the centre of the core. \\

Loading atoms inside a fibre is nontrivial and, to overcome some of the obvious limitations, Balykin \textit{et al.} \cite{Balykin04} proposed trapping $^{133}$Cs atoms outside the ONF by propagating a red-detuned laser beam in the fundamental guided mode (i.e. HE$_{11}$). This beam creates a dipole force gradient within the evanescent field which attracts the surrounding atoms towards the fibre.\\

\begin{figure*}
 \centering
        \includegraphics[width=0.8\textwidth]{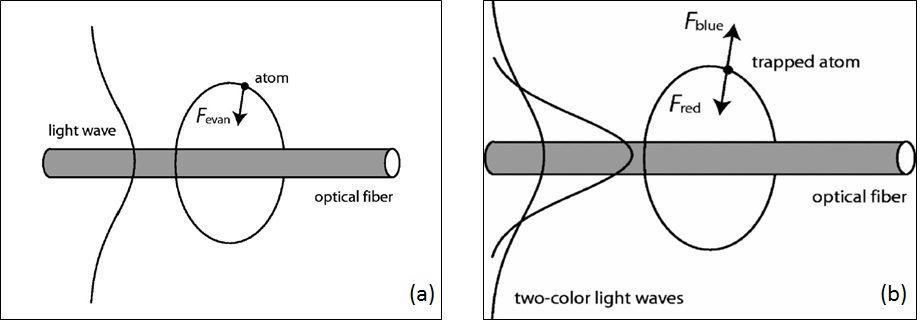}
  \caption{(a) Single colour trap involving only a red-detuned beam to counter-balance the centrifugal force of atoms initially revolving around the ONF. Image taken from \cite{Balykin04}. (b) Two-colour trap involving a red-detuned beam to attract atoms and a blue-detuned beam to prevent them from touching the ONF's surface. Reproduced from \cite{Lekien04}.}
  \label{fig:firstONF}
 \end{figure*}

To understand the forces at play, consider a single atom placed in an electric field, \textbf{E}(\textbf{r}). This electric field induces a dipole moment $-e\textbf{r}$ on the atom such that \cite{book:Foot13}
\begin{equation}
-e\textbf{r} = \epsilon_0\chi_a\textbf{E}(\textbf{r}),
\end{equation}
 where $e$ is the fundamental electric charge, $\epsilon_0$ is the electric permittivity in vacuum and  $\chi_a$ is the scalar polarisability. The interaction between this dipole and the electric field yields a potential energy, U(\textbf{r}), given by
\begin{equation}
\text{U}(\textbf{r}) = \frac{1}{2}e\textbf{r}\:\textbf{.}\:\textbf{E}(\textbf{r}) ,
\end{equation}

\noindent where the factor of $1/2$ arises due to the dipole moment being induced rather than permanent.
Assume the electric field is a plane wave, propagating along the $z$-direction and polarised in the $x$-direction with angular frequency, $\omega$, and wave number, $k$, giving $\textbf{E}(\textbf{r}) = E_0(\textbf{r})\cos(\omega t - kz) \hat{\textbf{e}}_x$. Then, the force acting on the atom in the $z$-direction is expressed as
\begin{equation}
\label{eq:dipole}
\begin{aligned}
& \text{F}_z  = \frac{-\partial \text{U}}{\partial z} = -\frac{ex}{2} \frac{\partial E}{\partial z} \\
& = - \frac{ex}{2} \left[\frac{\partial E_0}{\partial z}\cos(\omega t - kz) + kE_0 \sin(\omega t - kz)\right].
\end{aligned}
\end{equation}
The expression in Eq.\eqref{eq:dipole} can be written with the dipole moment expressed in terms of its components in the Bloch sphere and time averaged to get the final expression (see for example p.198 of \cite{book:Foot13} for a detailed development)
 \begin{equation}
 \label{eq:dipole2}
   \begin{aligned}
 \bar{\text{F}}_z = & - \frac{\hbar \delta}{2} \frac{\Omega}{\delta^2 + \Omega^2 /2 + \Gamma^2 / 4} \frac{\partial \Omega}{\partial z} \\  & +  \frac{\hbar k \Gamma}{2} \frac{\Omega^2 /2}{\delta^2 + \Omega^2 /2 + \Gamma^2 / 4},
 \end{aligned}
 \end{equation}
 where $\Gamma$ is the natural width of the excited state, $\Omega$ is the Rabi frequency and $\delta = \omega_L - \omega_0$ is the detuning between the laser frequency, $\omega_L$, and the resonant frequency, $\omega_0$, of the considered transition. The first term of Eq. \eqref{eq:dipole2} describes the dipole force, whereas the second term is the scattering force playing a central role in the initial laser cooling process. It is important to note that the dipole force depends on the gradient of the field intensity and the detuning. For negative detuning (i.e. a red-detuned beam), an atom placed in the vicinity of the ONF and interacting with its evanescent field will experience a force pointing in the same direction as the gradient of intensity, i.e. towards the ONF's centre. For fibres with a waist diameter roughly half the light's wavelength, this force can be balanced by the centrifugal force acting on atoms revolving around the ONF and, hence, can be used to trap them. However, the number of atoms that can be trapped using this scheme is restricted to those possessing an initial angular momentum along the fibre axis within an appropriate range. Le Kien \textit{et al.} \cite{Lekien04} proposed an improved version of the ONF trap, relying on a two-colour scheme; a red-detuned beam to attract the atoms towards the ONF and a blue-detuned beam to repel and prevent them from sticking to its surface. In this case, the van der Waals potential created by the fibre \cite{Minogin2010,Frawley12} acts together with the red-detuned beam to attract the atoms, while the blue-detuned beam prevents them from hitting its surface. Both proposed traps are shown in Fig.\ref{fig:firstONF}.\\

\begin{figure}[h!]
 \centering
        \includegraphics[width=0.49\textwidth]{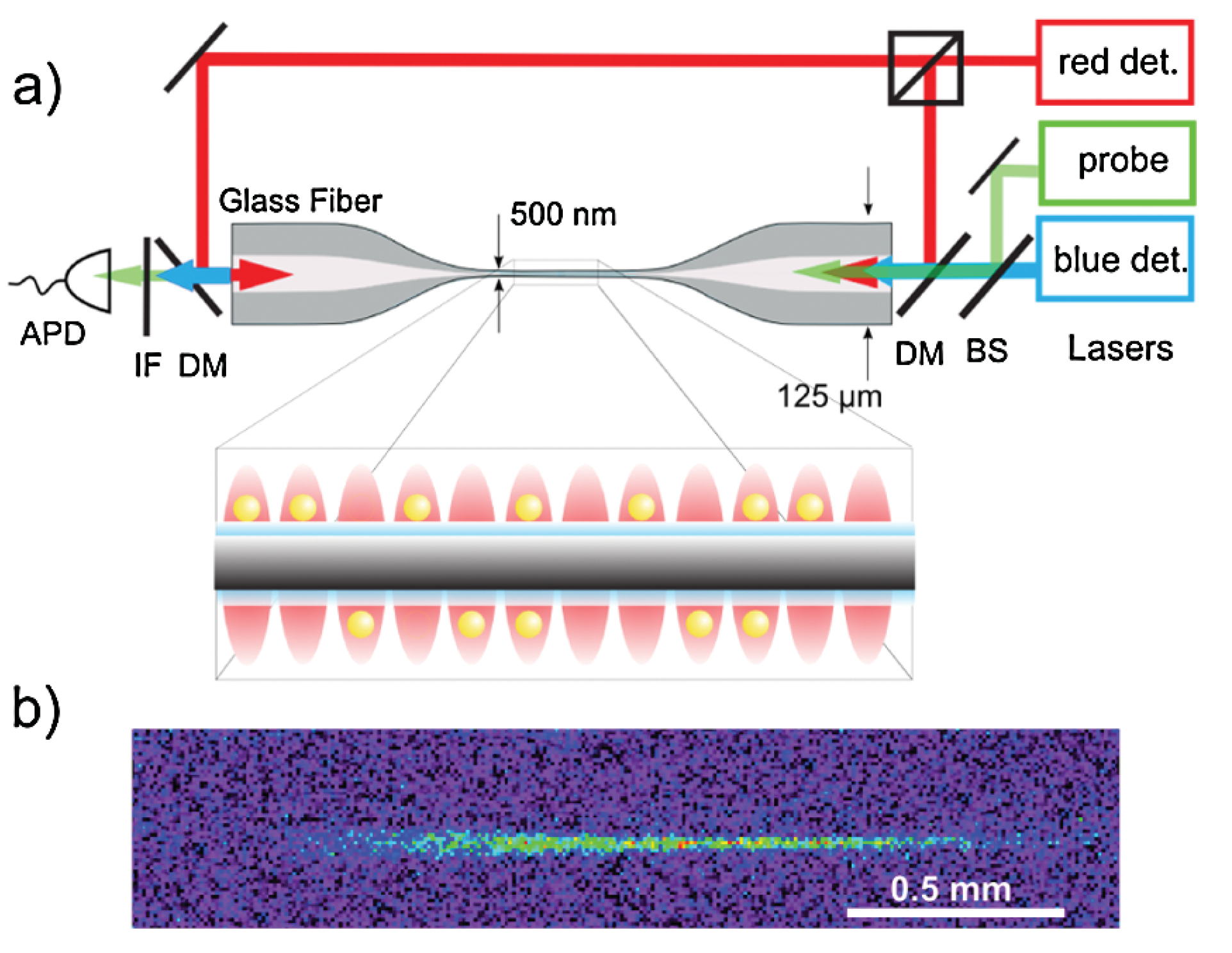}
  \caption{a) Layout of the experimental set-up and schematic of the standing wave trapping sites. b) Fluorescence image of trapped $^{133}$Cs atoms in the evanescent field of an ONF. Reproduced from \cite{Vetsch10}.}
  \label{fig:Arnotrap}
 \end{figure}

The experimental achievement of the two-colour trapping scheme was made in 2010 \cite{Vetsch10}. Laser-cooled $^{133}$Cs atoms were trapped around the waist of an ONF using the evanescent field resulting from the combination of a blue-detuned beam and two counter-propagating red-detuned beams with respect to the $^{133}$Cs D$_2$ line. The counter-propagating beams were used to generate periodically-spaced potential wells. The red- and blue-detuned beams were both linearly polarised, but orthogonal to each other. This gives better azimuthally localised trapping sites, as opposed to a parallel configuration. In addition, the trapping scheme was done in a collisional-blockade regime \cite{Schlosser02}, which allows only one atom at most per trapping site at a time. A weak probe beam, whose frequency was scanned around the $^{133}$Cs D$_2$ line, was launched into the fibre to generate fluorescence from the trapped atoms for imaging purposes (Fig.\ref{fig:Arnotrap}) and to provide an estimation of their number by saturated absorption. The resulting profile showed an asymmetry in the red-detuned region, which the authors attributed to a state-dependent light-shift created by the trapping beams. Subsequently, the group tuned their probe beam to the Stark-shifted resonant frequency and measured approximately 2000 atoms trapped around the ONF. Atoms trapped in this configuration showed ground state coherence times on the order of milliseconds \cite{Reitz13}. This trapping scheme was extended to create a conveyor belt \cite{Schneeweiss13} and led to the transportation of cold caesium atoms over a few mm along the ONF. Lee \textit{et al.} later adapted the two-colour scheme and trapped $^{87}$Rb atoms\cite{Lee14}. In contrast to the experiment in \cite{Vetsch10}, they arranged the trapping beams to have linear polarisations close to the parallel configuration. They justified this choice by the fact that the cold atom cloud had a better optical depth (OD) in this configuration, but were not able to explain the physical reason accounting for it.  An even more pronounced asymmetry in the absorption profile was observed and used to estimate the number of trapped atoms. Only 300 atoms were trapped in this experiment, that is 7 times less than reported in \cite{Vetsch10} for $^{133}$Cs atoms. This was attributed to the difference in atomic structure between $^{133}$Cs and  $^{87}$Rb, the latter being subject to larger vector light shifts. It was also proposed \cite{Schneeweiss14} to use the fictitious magnetic field generated by a nanofiber-guided light field in combination with an external magnetic bias field to create trapping potentials along the waist. Similar potential depths as in \cite{LeKien2013N} were reached, though in this work the power of the trapping field was an order of magnitude lower. In addition, this method allowed for better control over the trapped atoms' positions. \\

The above-mentioned light shifts can be problematic for the transfer of coherent information over long distances \cite{STILacroute12}. One solution is to use the so-called magic wavelengths \cite{STIkatori99}. When the trapping laser is operating at these specific wavelengths, hyperfine levels of interest undergo the same AC Stark shift, thus leaving the transition frequency unchanged. Several groups have been working on perfecting these state-insensitive traps, showing improved trap lifetimes and lower trapping powers in the two-colour scheme for both rubidium and caesium atoms \cite{STILekien05,STILacroute12,STIGoban12,STIArora12}. Although extremely efficient, state-insensitive traps are restrictive as one is required to work with a specific trapping wavelength.\\

\begin{figure*}
\centering
        \includegraphics[width=0.75\textwidth]{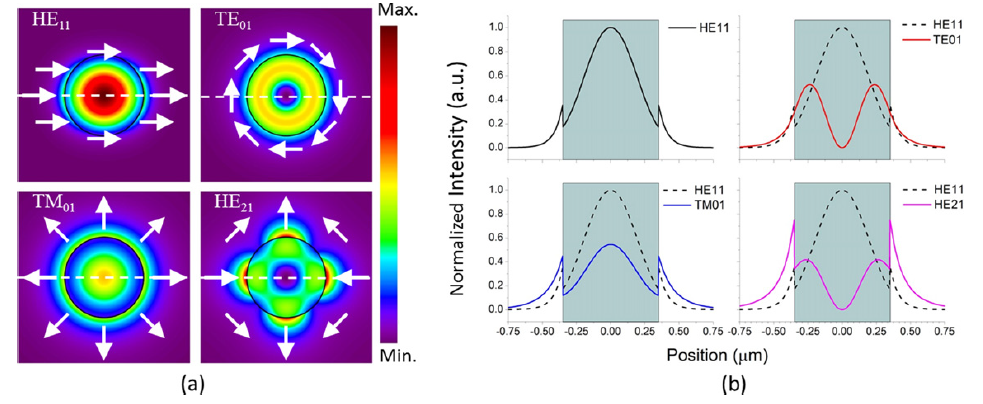}
  \caption{(a) Intensity profiles of the submodes in the first two fibre modes, i.e. LP$_{01}$ and LP$_{11}$. In each figure, the black circle denotes the cross-section of the fibre and the white arrows represent polarisation vectors. (b) Intensity profiles in each submode along the corresponding white dashed line in figure (a). The shaded zone represents the fibre and the dashed plots show the HE$_{11}$ mode for comparison. Reproduced from \cite{Kumar15HOM}.}
  \label{fig:RaviHOM}
\end{figure*}

HOMs offer a more flexible way to deal with the coherence issue. These modes have an extended and highly directional evanescent field at the waist compared to the fundamental mode. This is illustrated in Fig.\ref{fig:RaviHOM}, where the intensity profiles of the fundamental mode, HE$_{11}$, and the first set of HOMs, namely, TE$_{01}$, TM$_{01}$ and HE$_{21}$, are presented, along with their respective intensity distribution across the fibre waist. These properties could improve the coupling with laser-cooled atoms. Masalov and Minogin predicted that, if atoms were placed close to a fibre supporting HOMs, they would couple 5 to 10 times more light in these modes than in the fundamental mode \cite{Masalov13,Masalov14}. Our group demonstrated experimentally that, if one considers the ONF's LP$_{11}$ family of guided modes, the coupled fluorescence from laser-cooled $^{87}$Rb was six times more than if one only considers LP$_{01}$ \cite{Kumar15HOM}. The same experiment also showed that by injecting a Laguerre-Gaussian beam of index 1 (LG$_{01}$), i.e. a doughnut beam, to excite the LP$_{11}$ group of fibre modes, more light is absorbed by the  atoms surrounding the fibre, in accordance with the theoretical predictions in \cite{Masalov13}. We assumed this was due to the further extension of the evanescent field into the atom cloud. \\

\begin{figure}
 \centering
        \includegraphics[width=0.49\textwidth]{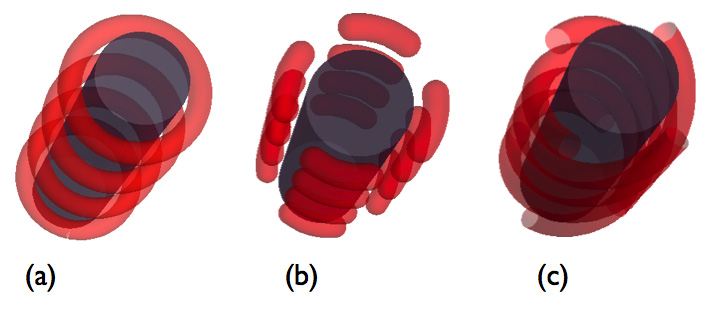}
  \caption{ Trapping geometries achievable for two counter-propagating HE$_{21}$ red-detuned beams.(a) Both beams have circular polarisation with the same handedness  yielding periodic ring-like trapping potentials. (b) Both beams have linear polarisation yielding periodic quadrupole-like potentials. (c) Both beams have circular polarisation but with opposite handedness yielding a double-helix trapping potential. Reproduced from \cite{Phelan13}.}
  \label{fig:helix}
 \end{figure}
 
Interference patterns resulting from the combination of two co-propagating blue-detuned LP$_{11}$ sub-modes, or one of the submodes and the fundamental mode, should yield deeper trapping potentials at lower powers \cite{Sague08} than that obtained relying solely on fundamental guided modes. Moreover, these modes offer the possibility to design trapping geometries that allow for better control over the atom trap positions. For example, the distance between each atom and their position relative to the fibre can be more accurately controlled by modifying the trapping beams' polarisations \cite{Reitz12,Phelan13}. This is depicted in Fig.\ref{fig:helix}, which shows the geometries proposed in \cite{Phelan13}. The two-colour scheme is used with counter-propagating red-detuned beams in a HOM and a blue-detuned beam in the fundamental mode. One can choose from a periodically spaced ring, a periodic quadrupole, or a double-helix trap geometry by choosing the red-detuned trapping beams' polarisations as circular-circular with same handedness, linear-linear, or circular-circular with opposite handedness, respectively. Fu \textit{et al.} also proposed to use a red-detuned beam in a superposition of both the TE$_{01}$ and HE$_{11}$ modes, in order to create a 1D optical lattice around the waist of an 820 nm diameter fiber \cite{Fu08}. To experimentally create these proposed traps, one has to be able to isolate each submode of the LP$_{11}$ group, namely TE$_{01}$, TM$_{01}$, and HE$_{21}$. This can be done either by injecting a free-space circularly polarised LG$_{01}$ beam into a few-mode fibre that supports the LP$_{11}$ group, and filter unwanted modes at the output via a half-wave plate \cite{Volpe04}, or by mixing two LG$_{01}$ beams with opposite topological charge (i.e. LG$_{01}$ and LG$_{0 -1}$) and opposite circular polarisation \cite{Maurer07}.\\

\subsection{Nonlinear optics in atomic media using ONFs}
\hspace{0.5cm}Among the numerous features offered by ONFs, one of the most interesting is their potential as tools for exploring nonlinear optics phenomena in atomic media. For example, it was proposed by Russell \textit{et al.} to use ONFs to provide intense light fields as needed for two-photon absorption by laser-cooled atoms \cite{RussellAIP}. This can be done using extremely low powers compared to similar experiments performed in free-space. Our group demonstrated this effect \cite{Kumar15AT,VandnaSPIE} by showing Autler-Townes splitting in laser-cooled $^{87}$Rb atoms via an ONF-mediated two-photon process. The Autler-Townes effect appeared for powers as low as 20 nW and frequency up-conversion was, in addition, observed with powers as low as 200 pW.\\
\begin{figure*}
\centering
        \includegraphics[width=0.8\textwidth]{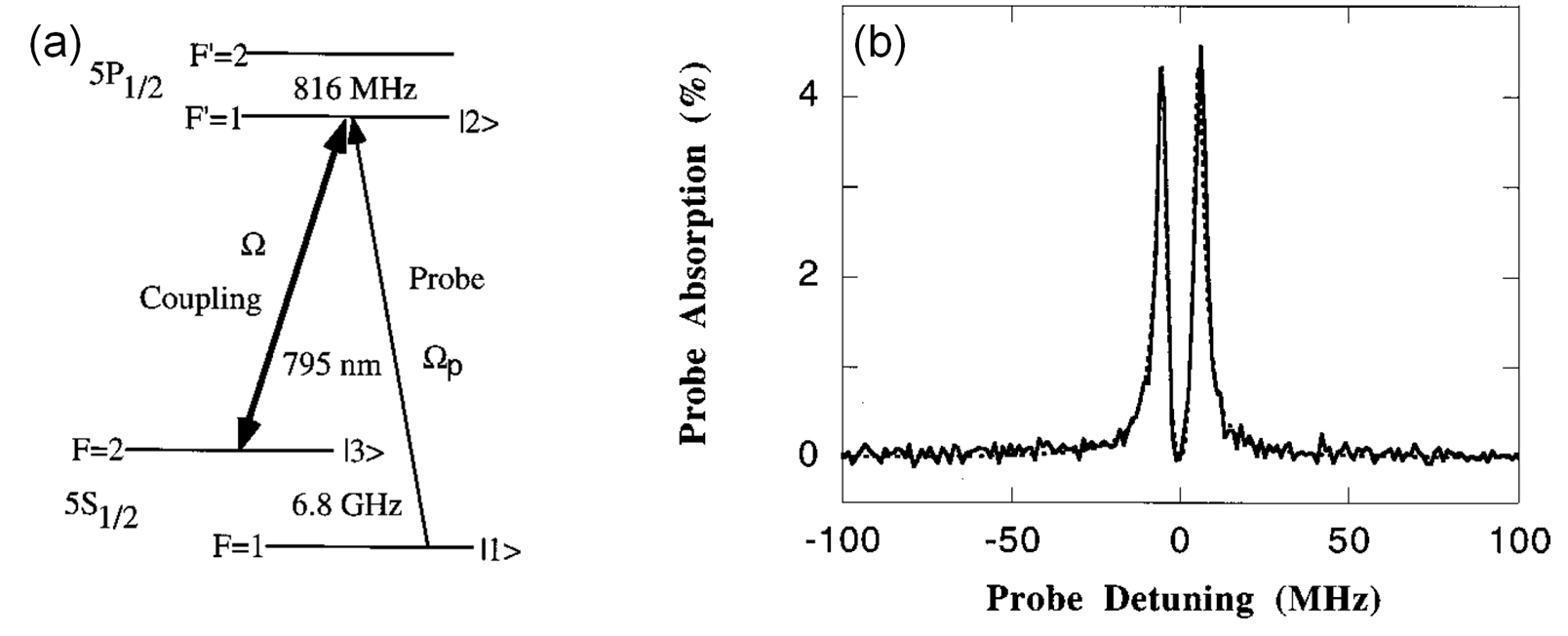}
  \caption{(a) Principle of the $\Lambda$-configuration EIT in $^{87}$Rb atoms, and (b) experimental observation of the EIT window when scanning the frequency of the probe beam. Reproduced from \cite{Yan01}.}
  \label{fig:gamma_config}
\end{figure*}

The development of an efficient quantum memory is fundamental to the creation of any quantum-based computing system. In a quantum network, a quantum memory acts as a ``node" connecting quantum channels in which information is written, stored and read out \cite{Kimble08}. Quantum information is easily lost over long distances due to the imperfect nature of information-carrying media such as optical fibres. This problem can be resolved by setting quantum repeaters along a transmission channel \cite{DLCZ01}. These repeaters store information twice so that one half can be ``destroyed" by a local measurement, which in turn allows the other half to be teleported to a distant node. The role of quantum memory in such a device is therefore obviously important. Light-atom systems are good candidates for the above mentioned purposes and extensive work on the implementation of optical quantum memories, reviewed in \cite{Lvovsky09}, has been done to date. In particular, it has been proposed to store information, carried from light, in atoms via EIT \cite{EITproposal}.\\

The EIT process, which was first demonstrated in 1990 with strontium atoms \cite{Boller90}, relies, in its simplest form, on three atomic levels, namely two ground states and an excited level. This is the commonly used $\Lambda$-configuration (see Fig.\ref{fig:gamma_config}), which will be used for the description hereafter. Note that other configurations such as the ladder- or V-configuration have also been presented \cite{Fleischauer05}. In the $\Lambda$-configuration, the two ground states can only couple via the single excited state of the system. The first ground state, lower in energy than the second one, is coupled to the excited state via a weak probe field. The second ground state, however, is coupled to the excited state via a strong control field. Typical transmission profiles of the probe beam when the control field is switched off and on are shown in Fig.\ref{fig:gamma_config}. It appears clearly on these profiles that, for a definite range of frequencies, the optical medium is transparent to the probe field. Aside from this transparency window where absorption would normally be observed, EIT also features a drastic reduction in the group velocity of the light beam passing through the material \cite{Hau99}.\\

In 2000, Lukin and  Imamo$\breve{\text{g}}$lu \cite{Lukin00} proposed to take advantage of this feature to create highly entangled photons. The main idea was to drive an EIT process in  $^{85}$Rb and $^{87}$Rb so that one isotope would serve as a slowing medium for a weak field, which in turn would be used to generate EIT in the second isotope. This work led to the seminal paper by Duan, Lukin, Cirac and Zoller describing the DLCZ protocol, at the heart of quantum repeaters \cite{DLCZ01}. As the phenomenon can be observed with a lesser contribution from Doppler broadening, EIT in cold atomic systems was also studied \cite{Veldt97,Hopkins97,Cataliotti97,Chen98,Yan01,Wang03}. \\

\begin{figure}
\centering
        \includegraphics[width=0.49\textwidth]{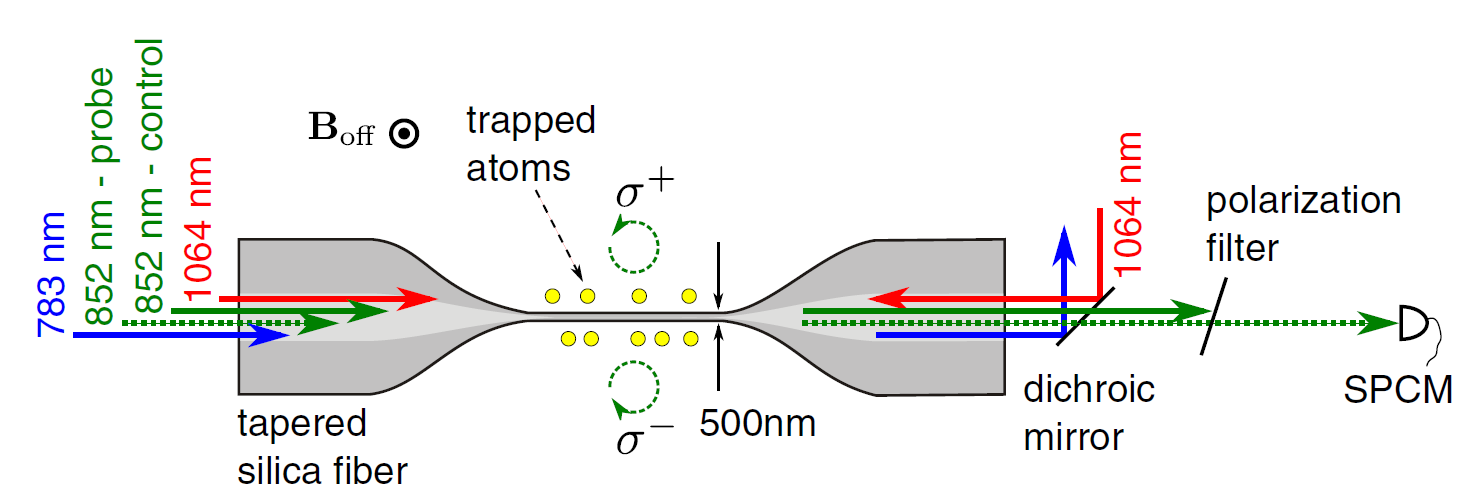}
  \caption{Experimental layout used in \cite{Sayrin15}. The red and blue arrows represent the trapping beams of the two-color scheme, whereas the plain and dotted green arrows represent the control and probe beams, respectively. At the output of the fibre, only the probe beam passes through the dichroic mirror and reaches the single-photon counting module (SPCM). Reproduced from \cite{Sayrin15}.}
  \label{fig:memory}
\end{figure}

It is also possible to generate EIT, and the subsequent slow light feature, in an atomic medium using the evanescent field at the waist of an ONF \cite{Patnaik02,Kien09}. Our group recently demonstrated a ladder-type EIT process in laser-cooled $^{87}$Rb atoms \cite{KumarEIT}, using powers below 1 $\mu$W in each of the required beams. We used a weak 780 nm wavelength laser as our probe beam to excite the transition $\text{5S}_{1/2}\rightarrow \text{5P} _{3/2}$, and a strong 776 nm wavelength laser as our coupling beam to excite the transition $\text{5P}_{3/2}\rightarrow \text{5D} _{5/2}$. These beams were injected, counter-propagating, in a single-mode ONF, with orthogonal linear polarisations. In addition to the EIT phenomenon, we showed an all-fibred optical switch in this system. Although our probe beam was kept on during the entire duration of the experiment, modulating the coupling beam yielded similar modulation in the transmission profile of the probe beam. Storage of guided light in a cold atomic system undergoing EIT was also demonstrated recently \cite{Sayrin15,Gouraud15}. In \cite{Sayrin15}, fibre-trapped $^{133}$Cs atoms, using a scheme similar to that presented in \cite{Vetsch10}, were used. As explained earlier, this scheme creates an optical lattice localised close to the waist of the ONF. Additionally to the trapping beams, co-propagating probe and control beams were also launched in the ONF to create the EIT process (Fig.\ref{fig:memory}). This resulted in the probe pulse being stored and retrieved on demand. However, the recovery efficiency was limited to 3\%. In \cite{Gouraud15},  a retrieval efficiency as high as 10\% is reported. As opposed to the experiment in \cite{Sayrin15}, this experiment used an ensemble of cold atoms instead of trapped atoms in a fibre-based optical lattice. Another difference was that only the probe beam was guided through the fibre while the control beams were in free-space.\\
\subsection{Fibres and cavities}
\begin{figure}[h!]
 \centering
        \includegraphics[width=0.48\textwidth]{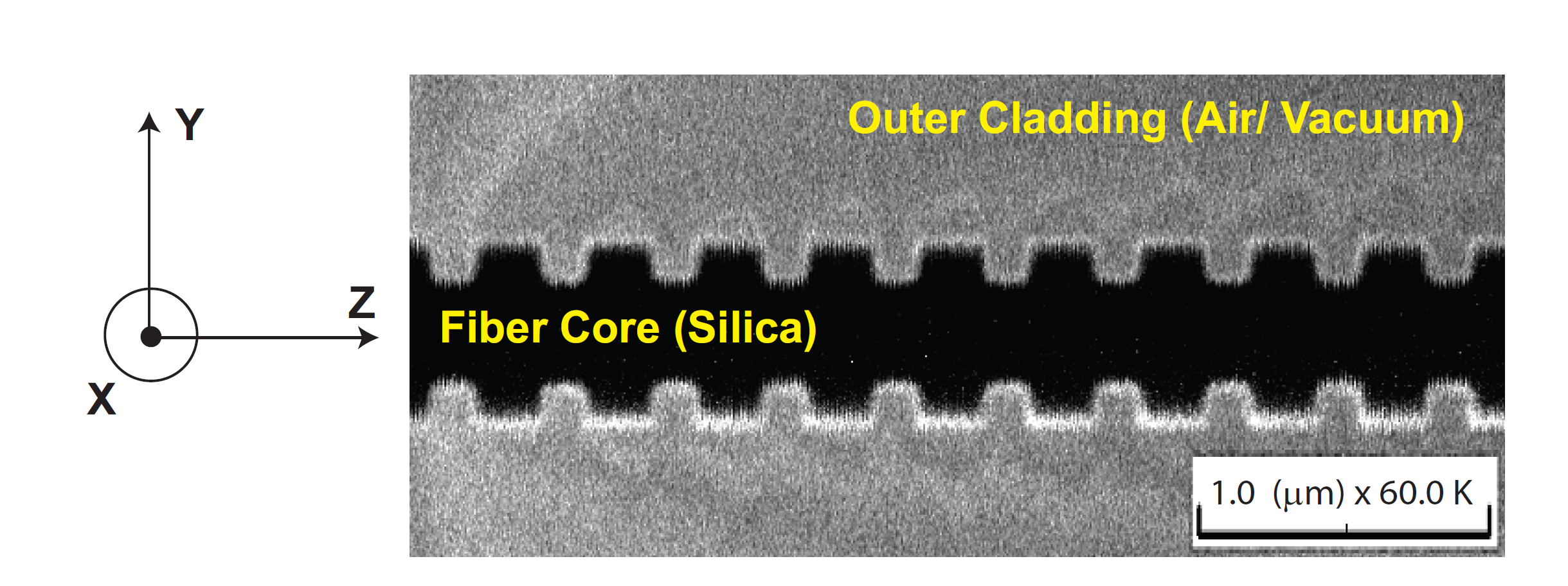}
  \caption{Scanning ion microscope image of a nanofibre Bragg grating fabricated via FIB milling. The fibre has a diameter of $\sim$560 nm and each slit is $\sim$150 nm wide and $\sim$100 nm deep. Two neighboring slits are separated by $\sim$360 nm. Reproduced from \cite{Nayak11}.}
  \label{fig:FBGfiber}
 \end{figure}

\hspace{0.5cm}Although ONFs offer similar features to those of cavities, albeit in a more scalable manner, there is an interest to combine both in order to improve the light and matter interactions. Some proposals have highlighted the advantages of creating a cavity network, in which each cavity is linked to others via fibres \cite{Busch2010,Kyoseva2012}. Le Kien \textit{et al.} have been studying in detail an alternative solution, consisting of ONFs with a fibre-Bragg-grating (FBG) positioned at either end of the waist \cite{cavLekien09,cavLekien092,LeKien2011}. The system combines both aspects in one and demonstrates strong coupling regimes with relatively low cavity finesse. Such devices could lead to improvements in the generation of EIT \cite{cavLekien09}, or entangled photons \cite{LeKien111}, and thus pave the way to quantum information generation and storage in an all-fibred network. These proposals are experimentally achievable as Nayak \textit{et al.} have reported fabrication of FBG-ONFs using a focussed ion beam (FIB) milling technique \cite{Nayak11} (see Fig.\ref{fig:FBGfiber}). Recently, a single caesium atom was trapped in a similar device \cite{Aoki2015} and strong coupling was achieved. The trap relied on a two-colour scheme analogous to that described in \cite{Vetsch10}, but with the red-detuned beams set at a magic wavelength in order to compensate for any light-shifts. As the atom was probed, it yielded a vacuum Rabi splitting, which is a signature of strong coupling between the cavity and the atom. Polarisation dependent FBGs can also be designed  \cite{Nayak13} by creating optical nanocavities at the surface of an ONF or OMF via femtosecond laser ablation.\\

\begin{figure}[h!]
 \centering
  \includegraphics[width=0.47\textwidth]{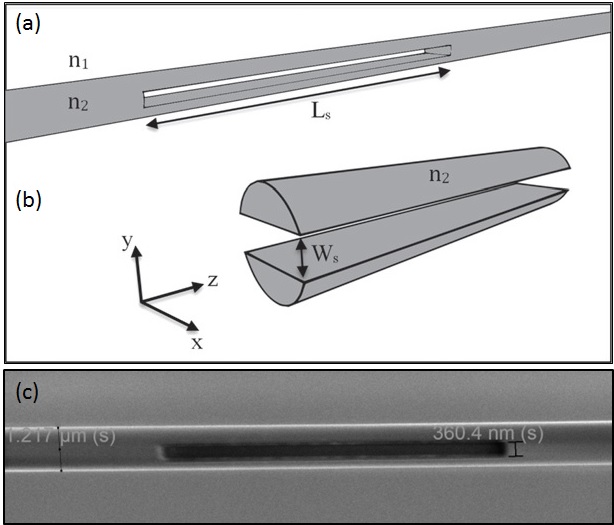}
  \caption{ Schematic representation of the slotted fibre proposed by our group. (a) Slot at the centre of the waist region, with L$_S$ representing the slot's length, and n$_1$ and n$_2$ the refractive indices of the fibre and the surrounding medium, respectively. (b) Cross-section of the slot in which W$_s$ represents the width of the slot. Reproduced from \cite{Mark14}. (c) SEM image of an actual STOF fabricated using the FIB milling technique.}
  \label{fig:slotfiber}
\end{figure}

Our group recently proposed a design consisting of a rectangular cavity at the centre of an OMF's waist\cite{Mark14}. This slotted tapered optical fibre (STOF), depicted in Fig.\ref{fig:slotfiber}, combines advantages from both the ONF and hollow-core fibres for atom trapping. The use of a two-colour trapping scheme in this device should lead to improved interactions between atoms and guided light compared to the case of atoms trapped outside the waist of an ONF. Short range interactions, such as van der Waals and Casimir-Polder effects, can easily be compensated by launching a blue-detuned beam in the STOF. The fabrication methods and achieved transmissions were recently reported \cite{Markspie}. To date, evidence of 100 nm colloidal particle trapping in such a device exists\cite{Markspie}.\\

\section{Conclusion}
\hspace{0.5cm}As presented in this review article, the large panel of features offered by ONFs makes them excellent tools for studies on cold atomic systems. Their potential for the creation of an atom-photon hybrid quantum system, with the advantage of it being fibred and thereby easily networked to other systems, is also clearly evident. A great amount of work is still to be done before achieving all-fibred networks. For example, a study on Rydberg atoms trapped next to ONFs \cite{KPS15} could lead to  applications in quantum information \cite{Brion12}. One could envision creating Rydberg atoms within a series of slots along the waist of an optical nanofibre \cite{Mark14} to form a 1D quantum network.  Alternatively, this system would provide an ideal testbed for exploring the quantum statistics of light in a strongly nonlinear atomic medium \cite{Grankin14}.\\

Likewise, to date very limited research has been conducted on HOM propagation in optical nano- or microfibres. Fibre-traps resulting from combinations of HOMs would lead to several distinct geometries and an overall better control over the trapped atoms' positions in the vicinity of the fibre \cite{Phelan13}. Moreover, the first set of HOMs, namely the LP$_{11}$ group of modes, can be easily excited by launching an LG$_{01}$ beam into the ONF. This beam carries orbital angular momentum (OAM) \cite{Allen92,Beijersbergen93} and provides extremely promising applications for atom trapping \cite{Kuga97,Okulov12} and quantum information processing \cite{Terriza07}. The details are beyond the scope of this review article and have been covered extensively elsewhere \cite{Arnold08,Yao11}. \\

Another promising aspect of the atom-fibre system worth investigating is the cooperativity between atoms trapped along an ONF since, due to the tight spatial confinement of the optical modes, the coupling strength per atom is increased when compared to free space beams. Taking advantage of this effect Qi \textit{et al.} estimated that a peak squeezing of about 5 dB should be attainable for a standard configuration consisting of 2500 atoms trapped 180 nm from an ONF with diameter 450 nm \cite{Qi15}.\\

In conclusion, while the field of ONFs, or even OMFs, integrated into atomic systems has reached a certain level of maturity in the last couple of years there are many more aspects that still remain to be explored and the full versatility of these ultrathin optical fibres in quantum engineering remains to be discovered in what promises to be an exciting field.

\section*{Conflicts of interest}
The authors declare no conflicts of interest.
\section*{References}
\providecommand{\newblock}{}

\bibliographystyle{iopart-num}
\end{document}